# Missing links towards understanding equilibrium shapes of hexagonal boron nitride: algorithm, hydrogen passivation, and temperature effects

*Jingzhao Zhang, Wenjing Zhao and Junyi Zhu*[*1]

There is a large discrepancy between the experimental observations and the theoretical predictions in the morphology of hexagonal boron nitride (h-BN) nanosheets. Theoretically-predicted hexagons terminated by armchair edges are not observed in experiments; and experimentally-observed triangles terminated by zigzag edges are found theoretically unstable. There are two key issues in theoretical investigations, namely, an efficient and accurate algorithm of absolute formation energy of h-BN edges, and a good understanding of the role of hydrogen passivation during h-BN growth. Here, we first proposed an efficient algorithm to calculate asymmetric edges with a self-consistent accuracy of about 0.0014 eV/Å. This method can also potentially serve as a standard approach for other two-dimensional (2D) compound materials. Then, by using this method, we discovered that only when edges are passivated by hydrogen atoms and temperature effects are taken into account can experimental morphology be explained. We further employed Wulff construction to obtain the equilibrium shapes of H-passivated h-BN nanosheets under its typical growth conditions at $T$ = 1300 K and $p$ = 1 bar, and found out that the equilibrium shapes are sensitive to hydrogen passivation and the growth conditions. Our results resolved long-standing discrepancies between experimental observations and theoretical analysis, explaining the thermodynamic driving force of the triangular, truncated triangular, and hexagonal shapes, and revealing the key role of hydrogen in h-BN growth. These discoveries and the advancement in algorithm may open the gateway towards the realization of 2D electronic and spintronic devices based on h-BN.

# Introduction

Hexagonal boron nitride (h-BN) is considered an important two-dimensional (2D) material because of its similar morphology with graphene and the wide band gap that may lead to important electronic device applications, such as 2D dielectric materials, spintronic materials and photovoltaic materials, etc.[1-5] However, the growth of high quality single crystalline h-BN with large areas has been challenging due to the difficulties of controlling its shapes, uniformity, and domain areas[6, 7] as well as high density of grain boundaries[8]. Therefore, in the past years, the morphology and growth mechanism of h-BN have attracted vast attention both experimentally and theoretically.

In experiments, the growth temperature of h-BN is usually high, ranging from approximately

[1] Department of physics, the Chinese University of Hong Kong, Shatin, N.T., Hong Kong, China. E-mail: jyzhu@phy.cuhk.edu.hk

Electronic supplementary information (ESI) available.

780 °C[9, 10] to 1065 °C.[7, 11] The growth methods for large-area films are mainly ambient-pressure chemical vapor deposition (APCVD)[7, 12] and low-pressure chemical vapor deposition (LPCVD).[13] Precursors like ammonia-borane ($H_3N-BH_3$)[8, 11] or borazine [$(HBNH)_3$][9, 10] are commonly used to grow BN thin film. Under such conditions, island shapes of h-BN with triangles, truncated triangles or hexagons have been observed frequently.[7, 11, 13-15]

It is extremely important to derive and obtain the correct equilibrium shapes of h-BN, as seen in experiments. Fixing the correct island shapes can be the first step towards the full understanding of its growth mechanism; such shapes can be obtained by Wulff construction[16] based on the absolute formation energies of the various types of edges. Therefore, it is important to have accurate estimations of absolute formation energies of these edges. Among the different types of edges, zigzag ones are the most theoretically challenging to calculate because of the asymmetric nature of them. Also, the bare zigzag edges are magnetized, with a net magnetic moment of one Bohr magneton at each edge atom.[4, 5] Therefore, to conduct an accurate simulation and stability analysis, the couplings among different bare edges should be forbidden. Also, spin-polarized calculations should be conducted.

In early literature, various approximation schemes to investigate edge stabilities were proposed. An average formation energy over two types of zigzag edges was obtained, but it was problematic to resolve the stabilities of each of them.[17, 18] Even for the bare-triangular cluster method with three equivalent edges, which is more recent and commonly used,[19-21] the unphysical charge transfer[22, 23] and the severe corner distortion[4] can largely hinder the computational accuracy for clusters with insufficient sizes. A recent investigation was based on clusters with large sizes, using the tight-binding density functional theory method,[20] but such big cells exceed the computational capability of typical first-principles calculations.[20] Though rough estimations, their qualitative conclusion that the zigzag edges should generally be less stable than armchair one still should hold;[17, 18, 21] however, the island edges observed in experiments are usually zigzag ones.[7, 9, 11] Therefore, there exist huge discrepancies between the theoretical and experimental results. There must be missing factors and/or physical mechanisms that may explain the unusual stabilities of the zigzag type of edges, which seem unstable at absolute zero from early formation enthalpy calculations.[17, 20, 21]

A possible explanation for the unusual stability is due to edge passivation. Passivation on surfaces or edges usually satisfy the electron counting model (ECM),[24] lowering the total energy.[25-27] One of the possible unavoidable impurities that passivate the edge is hydrogen (H). In the growth chamber, $H_2$ is usually a part of the carrier gas.[7] Also, the decomposition of precursors may produce extra H atoms.[8, 28-30] For example, it has been reported that $H_3N-BH_3$ would decompose into $H_2$, polyiminoborane (BHNH), and $(HBNH)_3$ at 130 °C.[28-30] Experimentally, $H_2$ gas was found to modify the shapes of h-BN film domains.[31] Also, the possible existence of N–H bonds of triangular rhombohedral-BN nanoplate has been proved by the Fourier transformation infrared spectra.[32] Although several investigations have been conducted on the electronic and magnetic properties of adsorbed H on top surfaces and edges of h-BN, as well as the stabilities thereof,[21, 33-35] the role of H-passivation of edges in h-BN growth, especially in influencing the morphologies of h-BN nanosheets, is still poorly understood.[19]

In addition, growth conditions that may affect the morphology of h-BN, such as temperature and pressure, have rarely been discussed in early literature. Temperature and pressure effects will tune the formation energies of H-passivated edges, mainly due to the vibrational entropy contribution of $H_2$ to the total free energy[36, 37]. This term is especially significant at high growth temperatures. Therefore, it is also necessary to consider such effects, which have been observed in compound semiconductors, such as GaN, ZnO, and the growth of their heterostructures,[27, 37, 38] in the thermodynamic analysis. Whether such effects can be exploited in h-BN growth is still an open question.

In this paper, we present an accurate method to calculate the absolute edge formation energy of 2D compound materials through a passivation-based scheme.[39, 40] Based on this method, we apply thermodynamic analysis[36-38] to investigate the effect of temperature and pressure on the morphology of H-passivated h-BN nanosheets. Then, by thermodynamic Wulff construction, we obtain the equilibrium shapes, which are consistent with experiments. The method in this paper can potentially serve as a standard approach applicable to other 2D materials as well as their lateral interfaces. Our results also show that temperature-stabilized H-passivation effect must be considered in determining the morphology of 2D nanosheets. While out of the scope of this paper, it should be noted that growth kinetics and substrate effects are also important aspects for understanding h-BN films growth,[19, 41] especially in the context of hydrogenation.

## Methodology

In previous wurtzite or zinc blende polar surface studies,[27, 39, 40, 42] a slab model with two conjugated polar surfaces was always applied, where one of the two surfaces (bottom surface) is passivated by correspondingly fractionally-charged pseudo-H atoms to satisfy ECM, avoiding unphysical charge transfers simultaneously. Later, by constructing pseudo-molecules or a set of tetrahedral clusters, the pseudo-chemical potentials of the pseudo-H atoms can be properly estimated, and the absolute formation energy can be obtained, to an accuracy of within several meV/Å$^2$.[39, 40] For 2D materials, no algorithms with such high accuracy have been proposed to calculate the absolute formation energy of asymmetric, or polar, edges. Schematic illustrations of polar [B-terminated (resp. N-terminated) zigzag edge, ZZB (resp. ZZN)] and non-polar edges (armchair edge, ARM) are shown in Fig. 1(a). For the non-polar edge, the edge energy can be directly obtained through a symmetric ribbon model.[21] To calculate the absolute formation energies of the asymmetric B- or N-terminated zigzag edges (ZZB, ZZN), we constructed a ribbon with the bottom edge passivated by H atoms [Fig. 1(b)]; for example, here the ZZN edge is passivated, satisfying ECM and maintaining the semiconducting nature. Therefore, the formation energy of the other (ZZB) edge with arbitrary configurations can be calculated by

$$\gamma_{\text{edge}} = \frac{1}{l}\left(E_{\text{tot}} - n_N\mu_N - n_B\mu_B - n_{H_N}\hat{\mu}_{H_N} - \sum_i n_i\mu_i\right). \tag{1}$$

Here, $l$ is the length of the edge, $E_{\text{tot}}$ is the total energy of this ribbon model, $n_N$ (resp. $n_B$) is the number of N (resp. B) atoms in this ribbon, and $\mu_N$ (resp. $\mu_B$) is the chemical potential of N (resp. B). In addition, $n_{H_N}$ is the number of H atoms used to passivate the bottom ZZN edge, and

$\hat{\mu}_{H_N}$ is the pseudo-chemical potential of H atoms, which consists of the chemical potential of H atoms and the energy of the passivated bottom edge, evenly distributed on each H atom. (The definitions and physical meanings of $n_{H_B}$ and $\hat{\mu}_{H_B}$, found below, are analogous.) Finally, $n_i$ is the number of $i$-type (say, H) atoms possibly adsorbed on the edge, and $\mu_i$ is the chemical potential of the said atoms. If we then assume a thermodynamic equilibrium between the bulk h-BN material and edges, we can obtain

$$\mu_N + \mu_B = E_{h\text{-}BN} = E_B + E_N + \Delta H_{h\text{-}BN}. \qquad (2)$$

Here, $E_{h\text{-}BN}$, $E_B$, and $E_N$ are respectively the total energy per formula of bulk h-BN material, and the total energies per atom of solid B and gaseous N$_2$; $\Delta H_{h\text{-}BN}$ is the formation enthalpy of h-BN. Therefore, Eq. (2) gives the constraint for $\mu_N$ and $\mu_B$ in Eq. (1), i.e. $E_j + \Delta H_{h\text{-}BN} \leq \mu_j \leq E_j$, where $j$ is B or N.

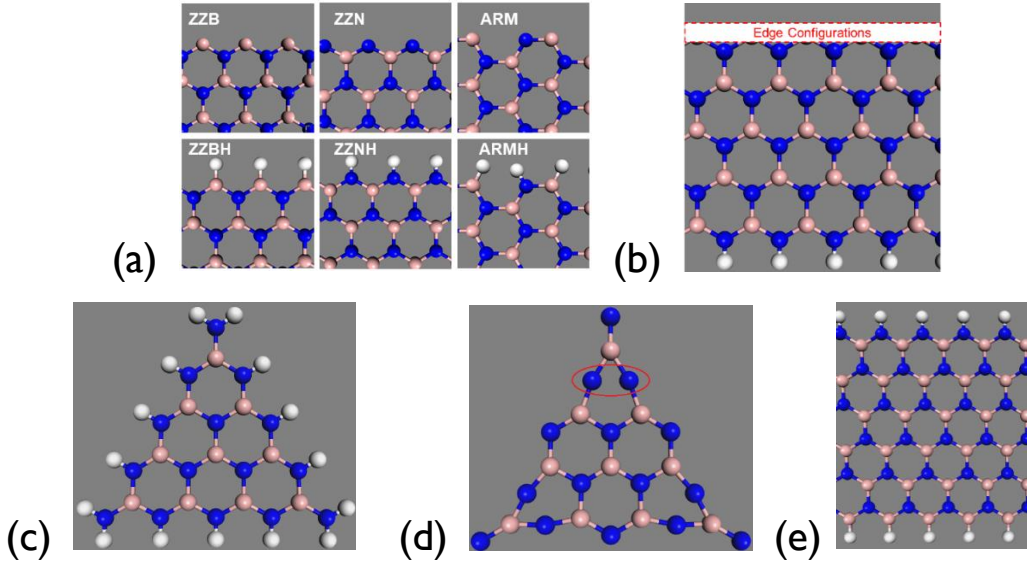

**Fig. 1** (a) Typical h-BN edge configurations with and without H-passivation. N, B and H atoms are denoted as blue, light pink and white spheres, respectively. (b) A ribbon model with one side fully-passivated for calculating asymmetric zigzag edges with any arbitrary configurations. (c) An example of a fully-passivated triangular cluster (with N termination and size $m = 5$). (d) An example of a bare triangular cluster (with N termination). The red ellipse shows that large distortions are induced at the corner. The two N atoms form bonds after relaxation, with the interatomic distance shrinking from ≈2.5 Å to ≈1.6 Å. (e) A ribbon models with two conjugated edges, both passivated by H atoms.

Next, drawing analogies between the three- and two-dimensional materials, we should be able obtain $\hat{\mu}_{H_N}$ through constructing a set of fully-passivated triangular clusters, similar to the one shown in Fig. 1(c). The total energy of such clusters can be expressed as

$$E_{\text{tot}}^{\text{cluster}} = \frac{m^2 + m}{2}\mu_N + \frac{m^2 - m}{2}\mu_B + (3m - 6)\hat{\mu}_{H_N} + 6\hat{\mu}_{H_N}^{\text{corner}}. \qquad (3)$$

Here, $m$ denotes the cluster size, using the number of atoms on the edge, which in the case of Fig. 1(c) equals five. $\hat{\mu}_{H_N}^{\text{corner}}$ is the pseudo-chemical potential of H at the corner of the cluster. By a non-linear fitting of Eq. (3), we are able to obtain the corresponding $\hat{\mu}_{H_N}$. $\hat{\mu}_{H_B}$ can be calculated by analogous treatments.

Since the sum of the pseudo-chemical potentials of the H atoms that passivate ZZN and ZZB is a constant, which can be obtained from a fully-passivated ribbon as shown in Fig. 1(e), it is instructive to check the sum of $\hat{\mu}_{H_N}$ and $\hat{\mu}_{H_B}$ obtained from different fully-passivated triangular clusters as a self-consistency check of our algorithm. Then, we can define a residual energy, quantifying the systematic error, as

$$E_\mathrm{r} = \frac{1}{2l}(E_\mathrm{p} - n_N\mu_N - n_B\mu_B - n_{H_N}\hat{\mu}_{H_N} - n_{H_B}\hat{\mu}_{H_B}), \qquad (4)$$

where $E_\mathrm{p}$ is the total energy of this ribbon. For a highly accurate algorithm, the residual energy should be close to zero.

Our total energy calculations were based on spin-polarized density functional theory (DFT)[43, 44] as implemented in VASP,[45] with a plane-wave basis set[46, 47] of an energy cutoff of 400 eV, using the PBE formulation of the generalized gradient approximation (GGA) of the exchange-correlation functional.[48] The ribbons of monolayer h-BN were separated by a vacuum of at least 15 Å. The k-point sampling is 13×1×1 for ribbons and gamma point-only for triangular clusters. The atoms in all the structures were allowed to relax until forces converged to less than 0.005 eV/Å. We have conducted careful convergence tests for all the aforementioned parameters and for the ribbon thicknesses.

# Results and Discussions

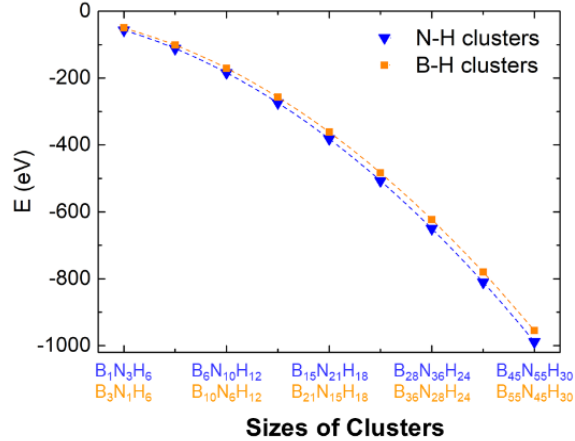

**Fig. 2** Total energies of the H-passivated clusters as a function of their sizes, with non-linear fittings.

To obtain the absolute formation energies of the edges, we need to first obtain the pseudo-chemical potentials of the H atoms that passivate the bottom edge of the ribbon. The pseudo-chemical potentials $\hat{\mu}_{H_N}$ and $\hat{\mu}_{H_B}$ were estimated by a non-linear fitting of the sets of fully-passivated triangular clusters with the sizes from $m = 2$ to $m = 10$ (Fig. 2). In this fitting, we treated the chemical potential of N and B (resp. $\mu_N$ and $\mu_B$) as variables because they may slightly deviate from the bulk values.[40] Therefore, by substituting the obtained $\hat{\mu}_{H_N}$ (or $\hat{\mu}_{H_B}$) into Eq. (1), we can obtain the absolute formation energies of the ZZN and ZZB edges, denoted as

$\gamma_N$ and $\gamma_B$ respectively (Table 1). The ZZN (resp. ZZB) edge is found to be more stable under N-rich (resp. B-rich) conditions. Here, we also plotted the absolute formation energies of ZZN, ZZB and ARM as a function of the chemical potential $\Delta\mu_N = \mu_N - E_N$ of N (Fig. 3, black lines). The ARM edge is always the most stable one throughout the whole chemical potential region. As such, we constructed the equilibrium shape corresponding to this case, which is an ARM-edged hexagon, as shown in Fig. 4(a). This indicates how under low growth temperatures, such hexagon shapes is expected to be thermodynamically stable, if we do not consider any impurity effects. However, it is experimentally established that under different chemical potential conditions, beside hexagons, other shapes like triangles can also exist. Moreover, only zigzag-terminated hexagons were observed by STM in experiment[9], despite how none of the previous theoretical works have found zigzag-terminated hexagons to be thermodynamically stable.

**Table 1** Calculated formation energies of h-BN edges with the passivation-based scheme and the bare-triangular cluster method respectively. The accuracy is presented in percentage format, by the ratio between the residual energy $E_r$, obtained through Eq. (4), and $[(\gamma_B + \gamma_N)/2]$.

|  |  | $\gamma_B$ (eV/Å) | $\gamma_N$ (eV/Å) | Accuracy (%) |
|---|---|---|---|---|
| **Passivation-based** | B-rich | 0.986 | 1.398 | 0.12 |
|  | N-rich | 1.358 | 1.027 |  |
| **Bare-triangular cluster** | B-rich | 0.985 | 1.320 | 3.43 |
|  | N-rich | 1.357 | 0.948 |  |

To compare the computational accuracy with existing algorithms, we also performed calculations by using the algorithm stated in early literature[19-21] using bare triangular clusters. To make fair comparisons, we use clusters of the same sizes in both sets of calculations. Results are shown in Table 1 above, and details are given in Supplementary Information. The formation energies of B-terminated edges are similar in both algorithms, because the edge morphology of the B-terminated bare triangles is similar to that of the ribbon. Meanwhile, for the N-terminated edges, large distortions are observed due to significant distortions in the corners; more detailed discussions are also shown in Supplementary Information. Here, we can use the residual energy $E_r$ calculated by Eq. (4) as an estimation of the systematic error of different algorithms; taking the ratio between $E_r$ and $(\gamma_B + \gamma_N)/2$, i.e. the average edge energy, allows us to quantitatively compare the general applicability of the algorithms. As shown in Table 1, the systematic error of our passivation-based algorithm is within 0.0014 eV/Å (less than 0.12%), which is one order of magnitude smaller than that of the bare-cluster method. Furthermore, in our tests using fully-passivated clusters of reduced sizes (from $m = 2$ to $m = 4$), we also obtained results with an acceptable systematic error of less than 0.003 eV/Å.

As discussed above and in the Supplementary Information, our method proposed here is significantly more accurate than the previous approaches. Besides, our method is also much more efficient, by virtue of (1) converging much faster because of minimal corner distortions; and (2) the generality of the pseudo-chemical potentials of the passivating H: once obtained, they are applicable to any type of edges by the ribbon model with one side properly passivated, needing no special designs of bare triangle clusters.

It should be noted that the energy difference between the ARM edge and the ZZB (or ZZN) edge is more than 0.2 eV/Å, much larger than the systematic error of our algorithm. *Therefore, the large discrepancy between the theoretical predictions and the experimental results cannot be attributed to algorithmic accuracy, and must rather be physically explained.*

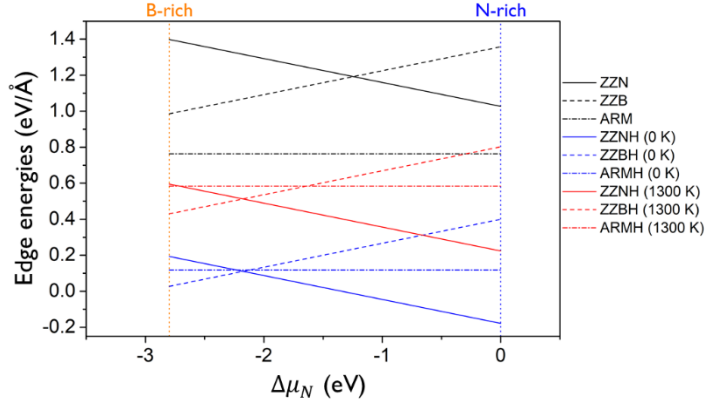

**Fig. 3** Absolute formation energies of different edges as a function of $\Delta\mu_N$, at different temperatures.

To explore the physical origin of the discrepancy, we must consider other factors, like impurities and growth conditions, such as temperature. In experiments, H atoms are widely distributed in the growth chamber. Since they are known to passivate the dangling bonds on the surfaces or edges to satisfy ECM and lower the overall energies, it is natural to investigate H-passivated edge as a first step to solve this puzzle.

Here, we calculated the absolute formation energy of H-passivated edges, i.e., ZZBH, ZZNH and ARMH (Fig. 3, blue lines). In this calculation, the chemical potential of H, $\mu_H$, is taken as half of the total energy of an $H_2$ molecule, which is obtained by absolute-zero DFT. The H-passivated edge configurations studied are shown in Fig. 1(a). Another type of configurations with two H atoms attaching to the same B (or N) atom[33] is also considered and calculated, but their energies are at least ≈1 eV/Å higher and are therefore not of interest. Compared with the formation energies of the bare edges, the formation energies of the H-passivated edges are lowered. The energy difference between B-terminated and N-terminated zigzag edges is increased due to the passivation and thus the N-terminated zigzag edge becomes more stable. As a result, the crossing point, where their energies are equal, shifts towards the N-poor limit (Fig. 3). The formation energy of ZZNH is negative in the region near the extreme N-rich limit. In this case, for h-BN crystal growth, a nanoribbon with the ZZNH edge is preferred thermodynamically. In addition, the energy gain for armchair edges is significantly smaller than that of the zigzag edges, because ECM is already satisfied for bare armchair edges, but not the bare zigzag edges, resulting in considerably smaller passivation effect. We then constructed the equilibrium shape at the crossing point ($\Delta\mu_N = -2.168$ eV; rf. Fig. 3), and it comprises ZZBH, ZZNH and ARMH in comparable ratios (≈2.53:2.53:1), as shown in Fig. 4(b); however, this shape was not observed experimentally either.

To form a complete understanding of the experimental morphologies, we have to consider the temperature effects. Experimentally, h-BN nanosheets are often formed at high growth temperatures. In this case, the entropy contribution of the gaseous $H_2$ phase also enters the formation energy formula. To estimate its contribution to the total energy, we expressed the H chemical potential as a function of temperature and pressure, by

$$\mu_H = \frac{1}{2}\left[E_{H_2} + 2\Delta\mu_H(T,p)\right], \quad (5)$$

where $E_{H_2}$ is the total energy of $H_2$ at absolute zero derived from first-principles calculations, and $\Delta\mu_H(T,p)$ captures the chemical potential relative to the total energy of the isolated molecule, determined by the gas atmospheric temperature and pressure. Theoretically, this energy term should include the contributions of the translational, rotational, and vibrational states,[36, 37, 49, 50] where the vibrational states of $H_2$ are dominant at high growth temperatures.[36, 37] Here, to estimate this energy term, following the method in our previous works,[27] we used the following expression for the quantity $\Delta\mu_H$:[27, 50]

$$2N\Delta\mu_H = G, \quad (6)$$

where $N$ is the number of $H_2$ molecules, and $G$ is the Gibbs free energy of gaseous $H_2$ in reference to absolute zero, which can be obtained from experimental data[51], considering the enthalpy and entropy contributions. This allows us to quantify $\mu_H$ via Eq. (5).

The proper temperature and pressure for evaluating the $H_2$ Gibbs free energy should be determined from the experimental conditions of h-BN growth. H-BN nanosheets of various shapes can be obtained by the APCVD technique at typical temperatures of 1000 °C to 1065 °C.[7] Therefore, we assumed typical growth conditions of $T$ = 1300 K, $p$ = 1 bar in determining the $H_2$ Gibbs free energy. It is noted that experimental observations of the equilibrium shapes with SEM, TEM, or other techniques are made at relatively lower temperatures, for example, around 200 °C for SEM,[7] raising potential concerns with the equilibrium shapes being affected by low-temperature thermal annealing after nanosheet formation; however, it was found that short periods of low-temperature annealing would not change the shapes significantly.[32] Therefore, we can still compare our computed equilibrium shapes with the experimental ones, which are characterized as grown.

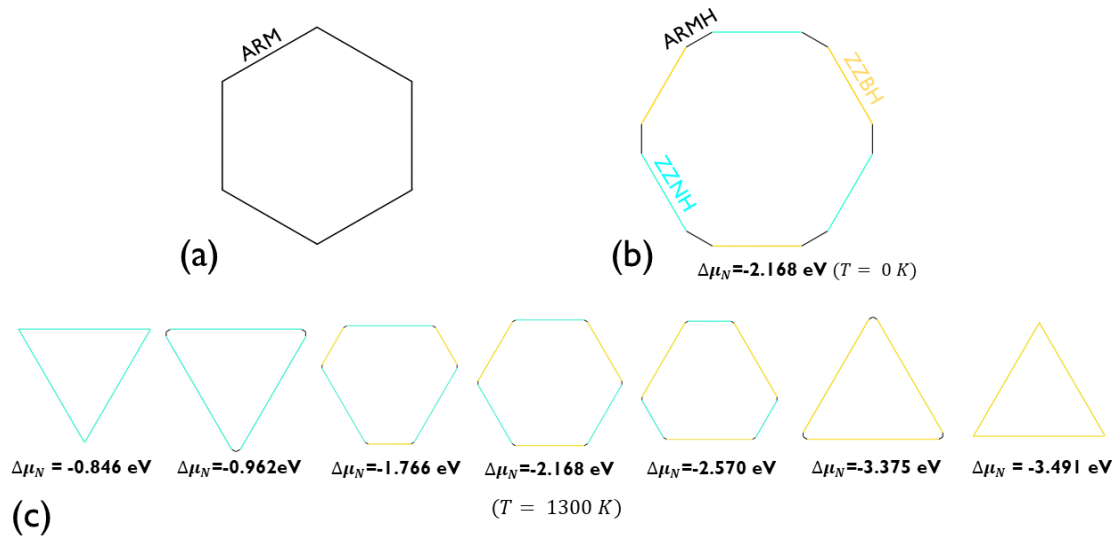

**Fig. 4** (a) A possible stable shape with bare edges. (b) An equilibrium shape with H-passivated edges at absolute zero. (c) Chemical potential-dependent equilibrium shapes consisting of H-passivated edges at 1300 K. All polar plots for the edge formation energies in the above situations are shown in Supplementary Information.

We calculated and plotted the absolute formation energies of the H-passivated edges, under the conditions $T$ = 1300 K, $p$ = 1 bar (Fig. 3, red lines). Compared with the absolute-zero case, the formation energy of all the edges shift upward. Also, the formation energy of the armchair edge increases more than that of the zigzag ones, due to the higher density of H on the armchair edge, incurring a higher energy penalty via the vibrational entropy of H. Therefore, it is possible to change the relative stability between the armchair and zigzag edges if we tune such temperature penalty.

Subsequently, we constructed the equilibrium shapes of h-BN nanosheets, as shown in Fig. 4(c). ZZNH-terminated triangles are found to be favored at the N-rich limit. When $\Delta\mu_N < -0.846$ eV, ARMH edges start to appear at the corners of the triangle; this shape we name as the armchair-corner triangle. When $\Delta\mu_N = -0.962$ eV, the ARMH edges compose about 10% of the whole perimeter. At this point, when the temperature is decreased, the composition of ARMH edge near the corner would slightly increase. These results are consistent with the experimental observations presented in Ref. [7], where it was described how the corners of the triangle become blunt when the growth temperature is lowered.

When $\Delta\mu_N < -0.962$ eV, ZZBH tends to co-exist with the other two types of edges, resulting in a truncated triangle; where, when $\Delta\mu_N = -1.766$ eV, the ratio between the ZZNH and ZZBH edge lengths becomes 2:1. Furthermore, at the crossing point at $\Delta\mu_N = -2.168$ eV (Fig. 3), a hexagonal shape would be favored, consisting of ZZNH and ZZBH edges with equal lengths, as well as a very small portion of ARMH edge, making up only ≈5.3% of the entire perimeter; it is difficult to resolve such short sections of ARMH edges from the existing micron-scaled SEM images.[7, 11] Also, because of the small ratio, ARMH may not appear at all in small nanosheets with perimeters shorter than ≈8.2 nm, where a primitive cell of the ARMH edge cannot form. In experiments, the observed deviation[11] of the hexagon angle from 120° may be attributable to the existence of such short ARMH edges.

As $\Delta\mu_N$ decreases continuously, an inverted shape transition process would happen; however, the edge energy difference between the ZZNH and ZZBH edges is much smaller than that in the N-rich region. As a result, even at the extreme B-rich limit, only truncated triangles may exist. This is also consistent with experiments, where only N-terminated triangles are observed.[7, 15, 52] In Ref. [7], the morphology of h-BN was found to be dependent on the chemical potential: as the N to B ratio decreases, the shapes are changed from triangular to hexagonal, with truncated triangular shapes in between. *Therefore, our obtained equilibrium shapes and chemical trend are consistent with the experimental observations.*[7, 11, 15]

Meanwhile, for the butterfly or star shapes,[6, 53-55] which incorporate concaved corners, other factors like substrate effects[19, 41] and growth kinetics[19, 56, 57] should also be considered; discussions of these, however, would be out of our present scope. Nevertheless, our algorithm, thermodynamic analysis, and equilibrium shape calculations lay a solid foundation for future works to further investigate these shapes and to fully understand the complex growth environment of BN.

## Conclusion

In contrast with early theoretical works, we discovered that most of the commonly seen equilibrium shapes of BN can be predicted and explained by a combination of H-passivation effect and the gaseous-$H_2$ entropy contribution. This important physical insight also suggests that it is highly possible to fine-tune the morphology of BN via the growth conditions, such as the partial pressures, the temperature, and the choice of chemical precursors. Based on these understandings, BN-based quantum dots with uniform sizes and controllable shapes may be grown.[58-61] Also, a similar strategy can be applied in the equilibrium-shape studies of other nanostructures or 2D materials.

In summary, we newly proposed an efficient and accurate computational method to calculate the asymmetric edges of h-BN nanosheets. The self-consistent accuracy is within 0.0014 eV/Å, much improved for more than an order of magnitude compared with previous methods. Combining this approach and thermodynamic analysis, we found that only by considering the H-passivation and temperature effects can theoretical predictions in agreement with experiments be made. For the first time, we calculated the equilibrium shape evolution as a function of the chemical potential, temperature, and pressure. As the growth condition changes from N-rich to B-rich, the island shapes undergo a gradual transition from N-terminated zigzag triangles, to armchair-corner triangles, then to ZZNH-dominant truncated triangles, hexagons, and finally to ZZNB-dominant truncated triangles. Our work presented not only significantly improves the computational accuracy, but also reveals the key role of H-passivation during the growth of h-BN nanosheets by thermodynamic analysis. Our computational method can potentially be adopted as standard practice towards the investigations of other important 2D materials, as well as their edges and interfaces.

# Conflicts of interest

There are no conflicts to declare.

# Acknowledgements

Thanks for the help of proof read and language polishing by Tsang Sze-chun. This work was supported by the start-up funding, HKRGC funding with the Project Code of 2130490, and direct grants with the Project Codes of 4053233, 4053134, and 3132748, at CUHK.

# Notes and References